\documentclass[12pt]{article}
\usepackage{amssymb}
\usepackage{amsmath}
\usepackage{amssymb}
\usepackage[dvips]{graphicx}
\date{}

\title{Quantum extension of European option pricing based on the Ornstein-Uhlenbeck process}
\author{Edward W. Piotrowski\\ Institute of Mathematics,
University of Bia\l ystok,\\ Lipowa 41, Pl 15424 Bia\l ystok,
Poland\\ e-mail: ep@alpha.uwb.edu.pl\\
 Ma\l gorzata Schroeder\\ Institute of Mathematics, University of Bia\l ystok, \\ Akademicka 2, Pl 15267 Bia\l ystok, Poland\\ e-mail: mszuli@math.uwb.edu.pl\\
 Anna Zambrzycka\\ Institute of Physics, University of Silesia, \\ Uniwersytecka
4, Pl 40007 Katowice, Poland \\ e-mail:
zanna12@wp.pl}
\begin{document}

\baselineskip9mm
\maketitle
\def\Z{{\bf Z\!\!Z}}
\def\R{{\bf I\!R}}
\def\N{{\bf I\!N}}
\begin{abstract}
In this work we propose a option pricing model based on the
Ornstein-Uhlenbeck process. It is a new look at the
Black-Scholes formula which is based on the quantum game theory.
We show the differences between a classical look which is
price changing by a Wiener process and the pricing is supported by
a quantum model.
\end{abstract}

PACS numbers: 02.50.-r, 02.50.Le, 03.67.-a, 03.65.Bz
 \vspace{5mm}
\newpage
\begin{flushleft}
{\Large\bf Introduction}
\end{flushleft}

Option trading experienced a gigantic growth with the creation of the Chicago Board of Options Exchange in 1973. 
Since this moment there is an on going demand for derivatives instruments. And
this induce financiers, mathematicians and physicians to the wider
and deeper research of the finance instruments of the price
changing dynamics. Owing to the fact, that an analogy has been
discovered between market prices behavior and the dust particle
motion model by a Wiener process. This observation revolutionized
derivatives pricing by developing a pioneering formula for
evaluating paying no dividend stock options, whose creators in
1997 have way a Nobel Prize for. The Black-Scholes formula
is the most popular deployed computational model. We propose
an alternative description of the time evolution of market price
making corrections in the Wiener-Bachelier model and which follows from the Ornstein-Uhlenback process.
This process has been successfully used by Vasicek in 1997 for
modeling the short time interest rate. Modifications have already been propose in works e.g. \cite{1,2}. 
In the latest year have
appeared variant a game theory based on the quantum formalism
\cite{3} it qualitatively broaden the capabilities of this
discipline describing the strategy which can not be realized in
classical models. Game theory describes conflict scenarios between a number
of individuals or groups who try to maximize their own profit, or
minimize profits made by their opponents. However, by adopting
quantum trading strategies it seems that players can make more sophisticated decisions, which may lead to better profit opportunities.
 The success of the quantum information theory (quantum algorithm or quantum
cryptography) could make these futuristic-sounding quantum trading
systems a reality, due to quantum computer development it will be
possible to better model the market. Quantum market model exist and in
this kind of market\footnote{quantum market can be created with using methods of the physics
which are used to make quantum effects. In this
kind of effect the rough consequences concerning a description of
reality since the century is shocking the explorers who are
faithful to the classical paradigm.} we can valuation the derivative
instrument. The pricing of derivative securities 
as a problem of quantum mechanics presented already Belal E.~Baaquie, see \cite{14}.  

In the first section we are going to quote the logarithmic equation which fulfills the logarithmic
stock price and we will be able to find the formula of these prices.
 In the second section we will be able to find the European option
 price supported by Wiener-Bachelier model and we will be able to
 receive the famous formula of Black-Scholes.
 Next we give the analogical
 probability like in the first section, but it will be supported
 by Ornstein-Uhlenbeck process.
 In the fourth section we interpret of the Ornstein-Uhlenbeck model in terms of 
 quantum market games theory as a non-unitary thermal tactic \cite{6}.
 In the fifth section with a little help of quantum model we find a model which describes the European option pricing.
\section{Price changing model as a Wiener process}
In 1900 Louis Bachelier in his dissertation showed that the
market prices behaves like a dust particles performing a
one-dimensional Brownian motion. Nowadays, we are used to approximate
the logarithmic price movements a Wiener process. These movements can be described by system of equations \cite{13}
$$
\left\{
\begin{array}{ll}
     \hbox{$dX_{t}=\sigma X_{t}\,dB_{t}+ \mu X_{t}dt$} \\
     \hbox{$X_{0}=x_{0}$} \\
\end{array}
\right. ,
$$
where $t$ equal time, $\mu $ is the growing rate (drift), $\sigma $
is a volatility rate and $B_{t}$ represent the standard Brownian motion. In the above 
process for the exact moment $t>0$, the
probability density function of finding the randomly moving
particle at the point $x$ (price logarithm) is given as
the Gaussian function
\begin{equation}
f(x,t):=\frac{1}{\sqrt{2\pi t}\,\sigma}\,\,\text{e}^{-\frac{(x-\mu
t)^{2}}{2\sigma ^{2}t}}.\label{f}
\end{equation}
Because the time evolution of the logarithm of stock price is described by
a Winer process, its current price $S_{t}$ is given by
$$
S_{t}=S_{0}\text{e}^{\mu t+\sigma B_{t}},
$$
where $S_{0}$ is the stock price in the $t=0$ moment. Assuming
that we are dealing with effective market and is no possibility
of arbitrage, and assuming that the parameters are fixed in this
model, the necessary estimators of these  parameters we can be found by econometric
methods. The mean value of the stock price at the moment $t$
 must equal $S_{0}$ and the
riskless interest rate $r$ is constant in the period $\left[ 0,t\right] $. So, if accepting the continuous
capitalization, we get
$$
S_{0}\text{e}^{rt}=\mathit{E}S_{t}=S_{0}\text{e}^{\mu
t}\mathit{E}\text{e}^{\sigma B_{t}}=S_{0}\text{e}^{\mu
t+\frac{1}{2}\sigma^{2}t}.
$$
Hence
\begin{equation}
\mu =r-\tfrac{\sigma^{2}}{2}\label{mu}\,.
\end{equation}
Substituting (\ref{mu}) to (\ref{f}) we are get distribution
of the logarithm of stock price
\begin{equation}
f(x,t)=\frac{1}{\sqrt{2\pi t}\,\sigma}\,\text{e}^{-\frac{(x-(r-\frac{1}{2}\sigma^{2})
t)^{2}}{2\sigma ^{2}t}} \label{fwb}.
\end{equation}
The above distribution is called the distribution of the
Wiener-Bachelier. The function (\ref{fwb}) fulfills a diffusion
equation with an arbitrage forbidding drift.
\section{The Wiener-Bachelier option pricing model}
Let us consider the European call option on company's
stock which is not paying a dividend, with the strike price of $K$ and
the maturity time $T$. Let us calculate the current value of this
option. The price of this option at the moment $T$ is equal
$[S_{T}-K]^{+}:= \max\{S_{T}-K,0\}$, because of $K\negthinspace
>\negthinspace S_{T}$ so this call option want be exercise and hence to that its value will be $0$ \cite{12}. 
The option fair price at
$T\negthinspace=\negthinspace 0$ should be equal to the discounted
mean value (assume that the interest rate $r$ is constant $[0,T]$
in the term)
\begin{equation}
c=\text{e}^{-rT}\mathit{E}[S_{T}-K]^{+}= \text{e}^{-rT}\int
_{-\infty}^{\infty}[S_{0}\text{e}^{x}-K]^{+}f(x,T)\,dx
\label{pierwsza}
\end{equation}
$$
=\text{e}^{-rT}\int
_{-\ln \frac{S_{0}}{K}}^{\infty}(S_{0}\text{e}^{x}-K)\,\frac{1}{\sqrt{2\pi T}\,\sigma}\,\text{e}^{-\tfrac{(x-(r-\frac{1}{2}\sigma^{2})T)^{2}}{2\sigma
^{2}T}}dx\,.
$$
Calculated of integral over the $dx$, we are getting the formula
$$
c=S_{0}N(\tfrac{\ln\frac{S_{0}}{K}+(r+\frac{\sigma
^{2}}{2})\,T}{\sigma
\sqrt{T}})-Ke^{-rT}N(\tfrac{\ln\frac{S_{0}}{K}+(r-\frac{\sigma
^{2}}{2})\,T}{\sigma \sqrt{T}})\,.
$$
In this way, we have received the famous Black-Scholes formula on the value
European call option, where the cumulative distribution function
$N$ is  the standard normal distribution.
\section{The model of Ornstein-Uhlenbeck}
The Ornstein-Uhlenbeck process describe a
particle behavior called by physicists the Rayleigh
particle \cite{4}. It can be described by the system of
equations
$$
\left\{
\begin{array}{ll}
     \hbox{$dX_{t}=\sigma \sqrt{2q }\,dB_{t}- q X_{t}dt$} \\
     \hbox{$X_{0}=x_{0}$} \\
\end{array}
\right. ,
$$
where $q $ and $\sigma $ are some positive constant, whereas
$B_{t}$ means the Brownian motion. The transition probability
density of the random variable $X$ in the time $t$, which is a
fundamental solution of the corresponding the Fokker-Planck
equation \cite{5}, is given by the formula
\begin{equation}
f_q(x,t):=\frac{1}{\sqrt{2\pi (1-\text{e}^{-2q
t})}\,\sigma}\,\,\text{e}^{-\frac{(x-x_{0}\text{e}^{-q t})^{2}}{2\sigma
^{2}(1-\text{e}^{-2q t})}}\,.
\label{moja}
\end{equation}
The parameter $q $ modifies the time scale in which we can examine
the particle place. For very small values of $q $ we obtain the short
time scale $\gamma \negthinspace:=\negthinspace qt$ \cite{4} in
which the Ornstein-Uhlenbeck process, which describe the velocity
evolution of the Rayleigh particle, approximate to the Brownian
motion\footnote{in physical models $X$ is a variable
which corresponds the velocity of the particle.}.\\
If we assume that the logarithmic price of the stock behaves similarly to the Rayleigh particle and we also take effective
market into consideration, so we can count the logarithmic
distribution of an stock based on Ornstein-Uhlenbeck model. Let us
$x=\ln S_{t}$ for $t\negthinspace>\negthinspace0$.
Then, making calculations similar for the Brown particle, we get
$$ S_{0}\text{e}^{rqt}=\mathit{E}S_{t}=\int
_{-\infty}^{\infty}S_{0}\text{e}^{x}f_q(x,t)\,dx=S_{0}\text{e}^{(x_0+
\sigma^{2}\sinh q t)\,\text{e}^{-q t}},
$$
and
$$
x_{0}(t)=rqt\,\text{e}^{q t}-\sigma
^{2}\sinh q t\,.
$$
In result the probability density of the logarithm of stock price is
\begin{equation}
f_q(x,t)=\frac{1}{\sqrt{2\pi (1-\text{e}^{-2q t})}\,\sigma}\,\,\text{e}^{-\frac{(x - rqt +\sigma^2\text{e}^{-q t}\sinh q t)^2}{2\sigma ^{2}(1-\text{e}^{-2q
t})}}\,.
\label{osta}
\end{equation}
Let us see that, $f_{q}\rightarrow f$ near $t\rightarrow 0$, see
Appendix. While near $t\rightarrow \infty$ function $f_q(x-rqt,t)$
approach to Gauss factorization with the same parameters, which
corresponds the price equality which result from the market
expectations, so on the base of the fundamental analyze. From here
the Ornstein-Uhlenback process would give the realistic situations
of the market in the mezzo-scale, temporal, it means in the time
where there is not enough to approximate by a  Wiener-Bachelier
process, but so small that in the result of the processes of
economics and innovations, there has not been changed the market
expectations, so like the asymptotic state of equality on the
``new'' one.

\section{Quantum market games interpretation of Ornstein-Uhlenback model}
In the quantum game theory Ornstein-Uhlenback process has the
interpretation of non-unitary tactics \cite{6}, leading to a new
strategy\footnote{these strategies are the Hilbert's spaces
vectors.}. We call tactics characterized by a constant inclination of an abstract market
player (Rest of the World) to risk
$\mathit{E}(H(\mathcal{P},\mathcal{Q}))\negthinspace=\negthinspace
const$ and maximal entropy thermal tactics. $H$ is a Hamilton operator that
specifies the model, whereas $\mathcal{P},\mathcal{Q}$ are
Hermitian operators of supply and demand \cite{6,7}. These traders
adopt such tactics  that the resulting strategy form a ground
state of the risk inclination operator $
H(\mathcal{Y},\mathcal{Z})$\footnote{that is they minimal eigenvalue, see \cite{6}.}. Therefor, thermal tactics are
represented by an operator\footnote{the operator
$H(\mathcal{Y},\mathcal{Z})-\frac{1}{2}$ annihilates the minimal
risk strategy.}
$$
\mathcal{R}_{\gamma}=e^{-\gamma
(H(\mathcal{Y},\mathcal{Z})-\frac{1}{2})},
$$
where
$\mathcal{Y}\negthinspace=\negthinspace\frac{1}{\sqrt{2}}(\mathcal{P}\negthinspace-
\negthinspace\mathcal{Q})$ and
$\mathcal{Z}\negthinspace=\negthinspace\frac{1}{\sqrt{2}}(\mathcal{P}\negthinspace+\negthinspace\mathcal{Q})$.
The variable $\mathcal{Y}$ describes arithmetic mean deviation  of
the logarithm of price from its expectation value. Quantum strategies create unique opportunities
for making profits during intervals shorter than the
characteristic thresholds for the Brownian particle. To
describe the evolution of the market price in the quantum models
we use the Rayleigh particle, it is the non-unitary thermal
tactics approaching equilibrium state\footnote{thermal tactics
lead asymptotically to the strategy with the smallest risk (the
ground state of the operator $ H(\mathcal{Y},\mathcal{Z})$ ),
concentrated around the price logarithm foretell in the fundamental
analysis with the market in balance remaining. Meanwhile after
the some time as a result of new information the market finds the
equality in the other part of the price logarithm, so it shifts
its ground state and as a effect its center wander as a Brown's
particle.}.

The variable $y$ describes the logarithmic transactional price.
$\mathcal{R}_{\gamma}$ operator (tactics) transforms the strategy
$\langle y|\psi \rangle \negthinspace\in
\negthinspace\mathcal{L}^{2}$ and, in the $\gamma $ moment it has
the form  \cite{8}
$$
\mathcal{R}_\gamma(y,\psi ):=\langle y|\mathcal{R}_\gamma\psi\rangle=\int_{-\infty}^{\infty}
\negthinspace\negthinspace \mathcal{R}_\gamma(y,y') \langle
y'|\psi\rangle dy'\,,
$$
where
$$
\mathcal{R}_\gamma(y,y')=\tfrac{1}{\sqrt{\pi \sigma
(1-\text{e}^{-2\gamma})}}\,\,\text{e}^{{-\frac{y^2-
{y'}^2}{2\sigma }-\frac{(\text{e}^{-\gamma}y-y')^2}{\sigma
(1-\text{e}^{-2\gamma})}}}\,.
$$
$\mathcal{R}_\gamma$ tactics is called the Ornstein-Uhlenbeck process. Adoption of the 
thermal tactics means that traders have in view minimization of the risk within 
the available information on the market. So we adopt such a normalization of the operator of the tactics 
so that the resulting strategy is its fixed point. Conditions for the fixed 
point\footnote{parameter $q$ measures
rate at in which the market is achieves the strategy, which
is a fixed point of thermal tactics.} these tactics allows to interpret it as probabilistic
measure describe in the formula $(\ref{moja})$, see \cite{6}.
\section{Option pricing based on quantum model}
Let us move to the European call option pricing underlying on company's stock
which is not paying dividend and based on changing prices which
are not the Brownian motion but the Ornstein-Uhlenbeck process modeled. To
achieve the Black-Scholes formula, it is enough to count the
integral $(\ref{pierwsza})$ for the modified density of the
probability $f\negthinspace\to\negthinspace f_q$, assuring no
arbitrage definite formula $(\ref{osta})$. The formula for
the price of the option takes the form
\begin{equation}
c_q=\text{e}^{-rqT}\mathit{E}_q[S_{T}-K]^{+}= \text{e}^{-rqT}\int
_{-\infty}^{\infty}[S_{0}\text{e}^{x}-K]^{+}f_q(x,T)\,dx
\label{endna}
\end{equation}
$$
= \frac{1}{\sqrt{2\pi (1-\text{e}^{-2q T})}\,\sigma}\,\,\text{e}^{-rqT}\int
_{-\infty}^{\infty}[S_{0}\text{e}^{x}-K]^{+}\,\text{e}^{-\frac{(x - rqT +\sigma^2\text{e}^{-q T}\sinh q T)^2}{2\sigma ^{2}(1-\text{e}^{-2q
T})}}\,dx
$$
$$
=S_0 N(\tfrac{(rqT+\ln\tfrac{S_0}{K})\,\text{e}^{qt} \,+\,\sigma^2 \sinh qt}{\sigma\sqrt{2 \text{e}^{qt}\sinh qt}}\,)\,\,-\,\,\text{e}^{-rqT}K N(\tfrac{(rqT+\ln\tfrac{S_0}{K})\,\text{e}^{qt} \,-\,\sigma^2 \sinh qt}{\sigma\sqrt{2\text{e}^{qt}\sinh qt}}\,)\,.
$$
The difference between the logarithmic
price the European call option described by the Ornstein-Uhlenback
process and logarithmic price call option described by the
Wiener-Bachelier process is present in Figure 1. The corrections to the Bachelier
model should matter for mezzo-scale.
\begin{figure}
\begin{center}
\includegraphics[height=6cm, width=9cm]{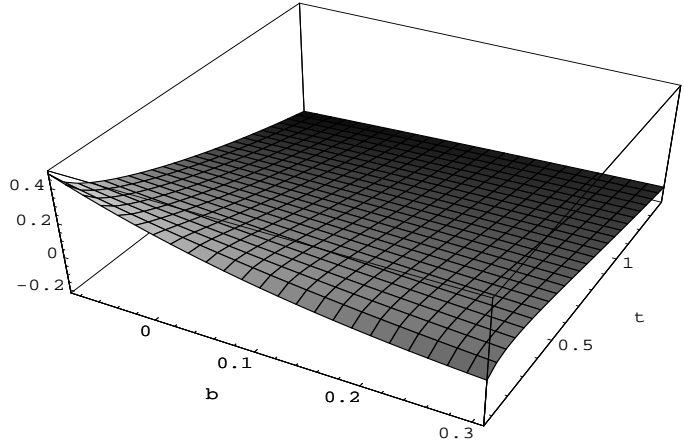}
\end{center}
\caption{Function $\ln c_{q}-\ln c$, for $b=\ln \frac{S_{0}}{K}$, $r=0$, $\sigma = q =1$\,.}
\end{figure}
\section{Final remarks}
We have proposed the alternative description of the time evolution
of market price that is inspired by quantum mechanical motion of
physical particles. Quantum market games broaden our horizons and
quantum strategies create unique opportunities for making profits
during intervals shorter than the characteristic thresholds for an
effective market (Brownian motion). On such market prices
correspond to Rayleigh particles approaching equilibrium state. Observations of the 
prices on the quantum market
can result quantum Zenno's effect \cite{7}, which should
broaden the range of correctness at the market description in the mezzo-scale. 
Sometimes it has side effects in the form of big jumps
like crashes in the market expectations in relation to new
asymptomatic equilibrium state.
 Quantum arbitrage based on such phenomena seems to be
feasible. The extra possibilities offered by quantum game
strategies can lead to more successful outcomes than purely
classical ones. This has far-reaching consequences for trading
behavior and could lead to fascinating effects in
quantum-designed financial markets. Quantum market games suggest
that such trading activity
 would take place on a ``quantum-board'' that contained the sets of all possible states of the trading game. However, 
 to implement such a game would require dramatic advances in technology, see \cite{11}.
 But it is possible that some quantum effect are already being observed. Let us quote the Editor's Note to Complexity
 Digest 2001.27(4) ``It might be that while observing the due ceremonial of everyday market transaction we are in fact
 observing capital flows resulting from quantum games eluding classical description. If human decisions can be traced to
 microscopic quantum events one would expect that ``Nature'' would have taken advantage of quantum computation in evolving complex brains.
 In that sense one could indeed say that quantum computers
 are playing their market games according to quantum rules.''
\section{Appendix}
Using the theorem about moments \cite{9}, the density of
probability is unambiguously definite by its cumulative moments.
The moment generating function for the Wiener-Bachelier density (\ref{fwb}), for the exact time $t=const$ is equal
$$
\Phi_{X}(\lambda)=\int _{-\infty}^{\infty}\text{e}^{\lambda
x}f(x)\,dx=\text{e}^{\frac{1}{2}\lambda (2r+(\lambda-1 )\sigma
^{2})t}\,.
$$
The first cumulative moment is
$$
s_{1}=(\ln \Phi_{X} )'\left|_{\lambda \rightarrow 0}=\Phi_{X}'(\lambda )\left|_{\lambda \rightarrow 0}=rt-\tfrac{\sigma^{2}}{2}t\, ,
\right.
\right.$$
the second cumulative moment is
$$
s_{2}=(\ln \Phi_{X} )''\left|_{\lambda \rightarrow 0}=\Phi_{X}''(\lambda )-(\Phi_{X}'(\lambda ))^{2}\left|_{\lambda \rightarrow 0}=\sigma ^{2}t\,,
\right.
\right.$$
the rest of the cumulative moments vanish. The first
cumulative moment measures the mean value and the second one measures
the risk.

For the density of the Ornstein-Uhlenbeck given by formula
(\ref{osta}) the generating function is
$$
\Phi_{X_{q}}(\lambda )=\int _{-\infty}^{\infty}\text{e}^{\lambda
x}f_{q}(x)\,dx=\text{e}^{\lambda rqt+\text{e}^{-qt}(\lambda-1
)\lambda \sigma ^{2}\sinh qt}\,,
$$
The first cumulative moment equals
$$
s_{1_{q}}=\Phi_{X_{q}}'(\lambda )\left|_{\lambda \rightarrow 0}=rqt-\sigma ^{2}\text{e}^{-qt}\sinh qt=rqt-\tfrac{\sigma^{2}}{2}(1-\text{e}^{-2qt})\,,
\right.
$$
The second cumulative moment equals
$$
s_{2_{q}}=\Phi_{X_{q}}''(\lambda )-(\Phi_{X_{q}}'(\lambda ))^{2}\left|_{\lambda \rightarrow 0}=\sigma ^{2}(1-\text{e}^{-2qt})\,,
\right.$$
whereas the others equal zero. As you can see the moment in both
densities differ by a non linear time modification.

The Taylor series expansion of $\ln \Phi_{X_{q}}$ is given by
$$
\ln \Phi_{X_{q}}=\ln \Phi_{X}+\left( (q-1)r+\sigma ^{2}(\lambda
-1) \left( q-\frac{1}{2}\right)\right) \lambda t+O[t^{2}]\,.
$$
For $t\rightarrow 0$ cumulative moments of the
Ornstein-Uhlenbeck and  Wiener-Bachelier process are equal, so from the
theorems of the moments we get $f_{q}\rightarrow f$.

\end{document}